\documentclass[english]{article}
\usepackage[utf8]{inputenc}
\usepackage[T1]{fontenc}
\usepackage{babel}
\usepackage{amsmath}
\usepackage{graphicx}
\usepackage{subcaption}
\usepackage{fancyhdr}

\usepackage{graphicx}
\begin{document}

\title{A Data-Efficient Deep Learning Based Smartphone Application For Detection Of Pulmonary Diseases Using Chest X-rays.}

\author{Hrithwik Shalu\textsuperscript{2}, Harikrishnan P\textsuperscript{2}, Akash Das\textsuperscript{3}, \\ Megdut Mandal\textsuperscript{4}, Harshavardhan M Sali\textsuperscript{2}, Juned Kadiwala\textsuperscript{1{*}}}

\maketitle
\thispagestyle{fancy}
\noindent
1. University of Cambridge     
\\
2. Indian Institute of Technology Madras 
\\
3. Indian Institute of Technology Patna
\\
4. Kalinga Institute of Industrial Technology Bhubaneswar
\\
{*}corresponding author

\begin{abstract}
This paper introduces a paradigm of smartphone application based disease diagnostics that may completely revolutionize the way healthcare services are being provided.  Although primarily aimed to assist the problems in rendering the healthcare services during the coronavirus pandemic, the model can also be extended to identify the exact disease that the patient is caught with from a broad spectrum of pulmonary diseases. The app inputs Chest X-Ray images captured from the mobile camera which is then relayed to the AI architecture in a cloud platform, and diagnoses the disease with state of the art accuracy. Doctors with a smartphone can leverage the application to save the considerable time that standard COVID-19 tests take for preliminary diagnosis. The scarcity of training data and class imbalance issues were effectively tackled in our approach by the use of Data Augmentation Generative Adversarial Network (DAGAN) and model architecture based as a Convolutional Siamese Network with attention mechanism. The backend model was tested for robustness using publicly available datasets under two different classification scenarios (Binary/Multiclass) with minimal and noisy data. The model achieved pinnacle testing accuracy of  99.30\% and 98.40\% on the two respective scenarios, making it completely reliable for its users. On top of that a semi-live training scenario was introduced, which helps improve the app performance over time as data accumulates. Overall, the problems of generalizability of complex models and data inefficiency is tackled through the model architecture. The app based setting with semi live training helps in ease of access to reliable healthcare in the society, as well as help in effective research of rare diseases in a minimal data setting.
\end{abstract}

\section*{Introduction}
The increasing adoption of electronic technologies is widely recognized as a critical strategy for making health care more cost-effective. Smartphone-based m-health applications have the potential to change many of the modern-day techniques of how healthcare services are delivered by enabling remote diagnosis [1], but it is yet to realize its fullest potential. There has been a paradigm shift in the research on medical sciences, and technologies like point-of-care diagnosis and analysis have developed with more custom-designed smartphone applications coming into prominence.
Due to the high rate of infection, with the total number of confirmed cases exceeding twenty million since its recent outbreak, COVID-19 was chosen as the initial disease target for us to study. With studies confirming that chest X-rays are irreplaceable in a preliminary screening of COVID-19, we started with chest X-rays as the tool to detect the presence of coronavirus (COVID-19) in the patients.[2]
Chest X-ray is the primary imaging technique that plays a pivotal role in disease diagnosis using medical imaging for any pulmonary disease. Classic machine learning models have been previously used for the auto-classification of digital chest images [3][4]. Reclaiming the advances of those fields to the benefit of clinical decision making and computer-aided systems using deep learning is becoming increasingly nontrivial as new data emerge[5][6][7], with Convolutional Neural Networks (CNNs) spearheading the medical imaging domain [8]. A key factor for the success of CNNs is its ability to learn distinct features automatically from domain-specific images, and the concept has been reinforced by transfer learning [9]. However, the process of learning distinct features by standard supervised learning using Convolutional Neural Networks can be computationally non-efficient and data expensive. The above methods become incapacitated when combined with a shortage of data.
Our approach represents a substantial conceptual advance over all other published methods by overcoming the problem of data scarcity  using a one-shot learning approach with the implementation of a Siamese Neural Network. Contrasting to its counterparts, our method has the added advantage of being more generalizable and handles extreme class imbalance with ease. We leverage open chest X-Ray datasets of COVID-19 and various other diseases that were publicly available (refer datasets section)[10]. Once a Siamese network has been tuned, it can capitalize on powerful discriminative features to generalize the predictive power of the network not just to new data, but to entirely new classes from unknown distributions[11][12]. Using a convolutional architecture, we can achieve reliable results that exceed those of other deep learning models with near state-of-the-art performance on one-shot classification tasks.
The world is being crippled by COVID-19, an acute resolved disease whose onset might result in death due to massive alveolar damage and progressive respiratory failure [13]. A robust and accurate automatic diagnosis of COVID-19 is vital for countries to prompt timely referral of the patient to quarantine, rapid intubation of severe cases in specialized hospitals, and ultimately curb the spread. The definitive test for SARS-CoV-2 is the real-time reverse transcriptase-polymerase chain reaction (RT-PCR) test. However, with sensitivity reported as low as 60-70\%[14] and as high as 95-97\%[15], a meta-analysis concluded the pooled sensitivity of RT-PCR to be 89\%[16]. These numbers point out false negatives to be a real clinical problem, and several negative tests might be required in a single case to be confident about excluding the disease[17]. A resource-constrained environment demands imaging for medical triage to be restricted to suspected COVID-19 patients who present moderate-severe clinical features and a high pretest probability of disease, and medical imaging done in an early phase might be feature deficient [18][19]. Although the cause of COVID-19 was quickly identified to be the SARS-CoV-2 virus, scientists are still working around the clock to fully understand the biology of the mutating virus and how it infects human cells[20]. All these calls for a robust pre-diagnosis method, which hopes to provide higher generalization, work efficiently with insufficient feature data, and tackles the problem of data scarcity. This is where our proposed method of  Data Augmentation Generative Adversarial Network (DAGAN) exploited by a Convolutional Siamese Neural Network with attention mechanism comes into the picture, exhibiting a state of the art accuracy and sensitivity.

\subsection{Generative Adversarial Networks}
Generative Adversarial Networks (GANs) are deep learning based generative models which take root from a game theoretic scenario where two networks compete against each other as an adversary. The constituent network models – a Generative Network and a Discriminative Network play a zero-sum game. GAN architecture paved way for sophisticated domain-specific data augmentation by treating an unsupervised problem as a supervised one, thus automatically training the generative model.

\begin{figure}[h!]
\includegraphics[width=12cm, height=6cm]{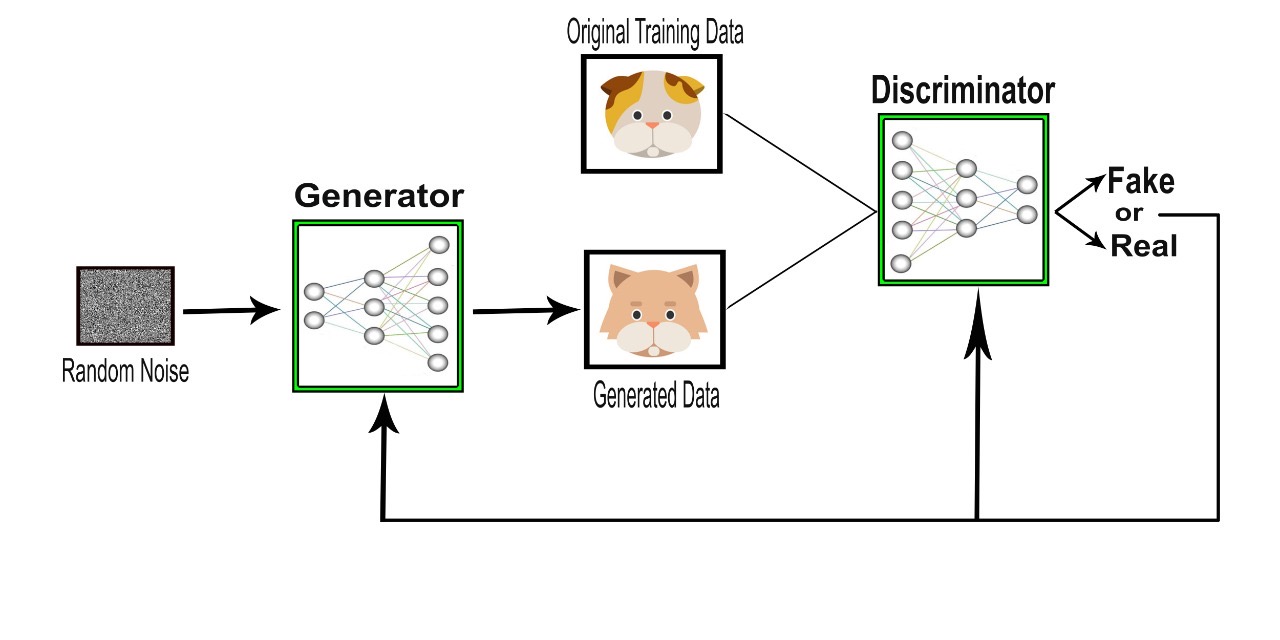}
\caption{Figurative representation of data flow in a Generative Adversarial Network}
\label{a1}
\end{figure}

The Generative Network utilizes the latent space, which is a projection or compression of the data distribution to generate plausible training examples. Latent space is the end result of mapping the points in the multidimensional vector space to points in the problem domain. Random vectors drawn from data distributions like the Gaussian distribution are used to seed the generative process. The Discriminative Network has the primary objective of classifying real (training set samples) and fake instances (generator samples). The Discriminative network component of the GANs are usually normal binary classification models, with real instances as positive examples and fake instances as negative examples. It connects to two loss functions, the generator loss and the discriminator loss. Training of discriminative networks penalizes the discriminator loss with the examples from generative networks with constant weights and real examples as its input, ignoring the generator loss. Similarly the generator is trained to create competent plausible samples and modifies its weights based on the generator loss.

\subsubsection{Data Augmentation Generative Adversarial Networks (DAGAN)}
Data Augmentation procedure is crucial in the training procedure of a deep learning model as it has proven to be an effective solution in tackling the problem of overfitting at numerous occasions. As the data could be made more generalized, by providing the same with suitable augmentation strategies. In the case of images, Augmentation plays a crucial role. As to correctly identify and recognise specific features in the same, a diverse set of considerably different sets of images are required. Image augmentation techniques are henceforth found in diversely different ways, ranging from simple transforms (rotation) to adversarial data multiplicative methods such the one we would be using for our purposes, called the data augmentation generative adversarial networks (DAGAN). 

\begin{figure}[h!]
\includegraphics[width=12cm, height=6cm]{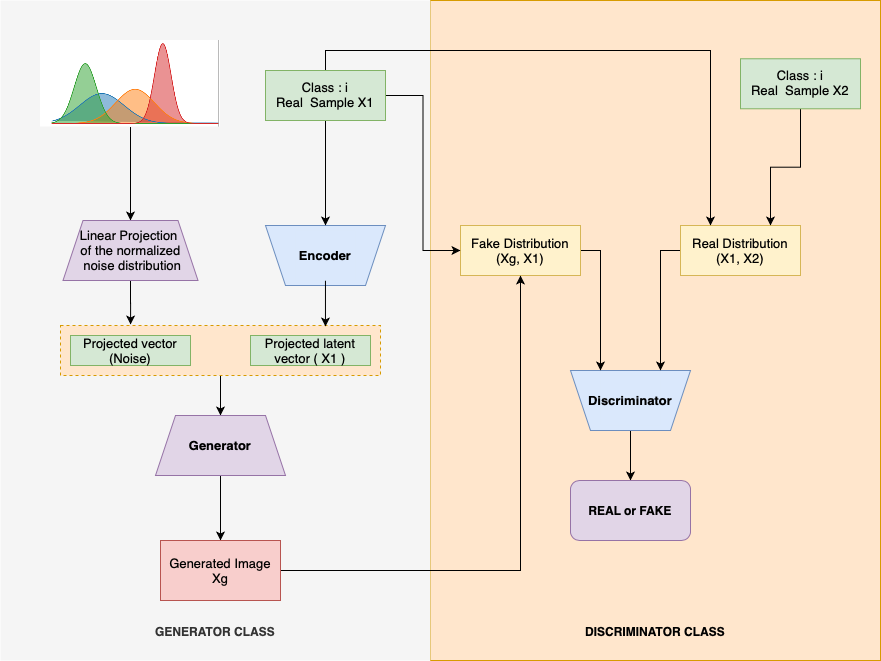}
\caption{The DAGAN architecture }
\label{a2}
\end{figure}

The purpose and uniqueness of DAGAN when compared to other types of GANs , is the ability to generate distinctive augmented images for any given image sample while preserving the distinctive class features intact. A general network architecture of the same is provided in Figure 2.

\textbf{Generator} : The Generator component of the DAGAN contains an encoder which provides a unique latent space representation for a given image and a decoder which generates an image given a latent representation. Any given image is first passed through the encoder to attain the corresponding latent representation, to which a scaled noise ( usually sampled from a Gaussian distribution ) is added to obtain a modified latent vector. The same is then passed through the decoder to obtain the corresponding augmented image.

\textbf{Discriminator} : The discriminator component of the DAGAN is similar to other GANs, where the basic purpose of which is to perform a binary classification to tell apart the generated and real images. The discriminator takes as input a fake distribution (generated images) and a real distribution (Images belonging to the same class).

\subsection{One Shot Learning and Siamese Networks}
Forming an ideal dataset for a typical multi-class classification task using standard supervised learning methods is quite difficult. In addition to class imbalance issues, data for certain tasks such as medical image analysis could rarely be collected to meet ideal standards. One-shot learning methods helps tackle these issues effectively. In the Deep Learning literature, Siamese Neural Networks are typically used to perform one-shot learning.

The Siamese Neural Network is a pair of neural networks trying to learn discriminative features from a pair of data points from two different classes.In our case the Siamese Networks would consist of two twin Convolutional Neural Networks which accept distinct inputs but are joined together by an energy function.The latent vector is the overall output from either of the twin neural networks, it is a unique and meaningful representation of individual images passed. In one shot learning the overall training objective is to obtain a vector valued function (Neural Network) which provides meaningful latent representation vectors to the each image passed. As any machine learning task, the one shot learning too has a loss function whose value conveys how close the network is in attaining optimal parameter values. In the case of Siamese Networks, the loss is a similarity measure between the latent vector outputs , enforced by a binary class label (like or unlike). 
The energy function takes as input the latent vectors formed by the CNN’s at their last dense layer (for each pair input passed) and outputs an energy value. The overall goal during the training process (optimization) can now be conveyed in terms of the energy function. The energy (output of the energy function) of a like pair is minimised and between unlike pairs it is maximised.

The typical energy functions used could be anything from a simple euclidean norm  to a fairly advanced function such as the contrastive energy function. A typical example of the contrastive energy is explained in brief.

The contrastive energy function takes in two vectors as input and in general performs the following computation.

\begin{equation}
\mathcal{L}(W) = \sum_{i=1}^{P}L(W, (Y, \vec{X_1}, \vec{X_2})^i) \label{eq1}
\end{equation}
Where,
\begin{equation}
L(W, (Y, \vec{X_1}, \vec{X_2})^i) = (1 - Y)L_S(D_W^i) + Y L_D(D_W^i)
\end{equation}
\indent
D\textsubscript{W} is the parameterized distance function as mentioned below 
\begin{equation}
D_W(\vec{X_1}, \vec{X_2}) = \left\|G_W(\vec{X_1}) - G_W(\vec{X_2}) \right\|_2
\end{equation}
\indent
Y is the binary class label. \\
\indent
L\textsubscript{S} and L\textsubscript{D} are functions chosen as per the task. \\

\begin{figure}[h!]
\includegraphics[width=12cm, height=6cm]{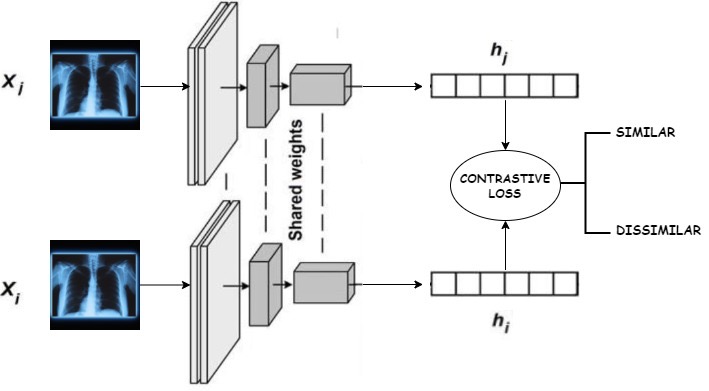}
\caption{ Feature comparison methods represented by a Siamese Neural Network architecture }
\label{a3}
\end{figure}

\subsection{Attention Mechanism}
Data no matter how clean will have irrelevant features, many of the predictive or analytic tasks does not rely on all of the features present in raw data. One of the factors that sets us humans apart from computers is our instinct of contextual relevance while performing any of our day to day activities. Our brains are adept at such tasks which makes us able to perform complex tasks quite easily. Attention is a deep learning technique designed to mimic this very property of our brain. Attention, as the name suggests is a methodology by which a neural network learns to selectively focus on relevant features and ignoring the rest. Attention was first introduced in the branch of natural language processing (NLP) [21], where it enabled contextual understanding for Sequence to Sequence models (Ex: Machine Translation) which led to better performance of the same. Attention mechanism in NLP solved the problem of vanishing gradients for Recurrent Neural Networks and at the same time brought in feature relevance understanding which boosts performance. The revolutionary impacts of deep learning paved way for creation of more efficient network architectures such as the Transformer (BERT) [22], which are widely applied these days. Moreover attention has been applied to other fields related to deep learning  such as the ones focusing on signal and visual processing.     

\subsubsection{Visual Attention Mechanism}
Images are a very abstract from of data, they contain numerous amounts of patterns (features) which could be analysed using latest computational tools to gain understanding on them. For many machine learning tasks such as regression or classification, identifying features of contextual relavance would improve the model performance and simplify the task. The same is the case for machine learning applied to images. For most images, the regions could be broadly classified as background and objects, where objects are of prime focus and background doesn't contribute to inference. Because of the same, knowing where to look and what regions to focus on while making an inference from images helps boost the performance of the model. Convolutional neural networks(CNN) are one of the best feature extraction tools for images in today's deep learning literature, attention applied to Convolutional features will help pick out relevant features of interest from the large pool of features extracted by a CNN.

\begin{figure}[h!]
\includegraphics[width=12cm, height=6cm]{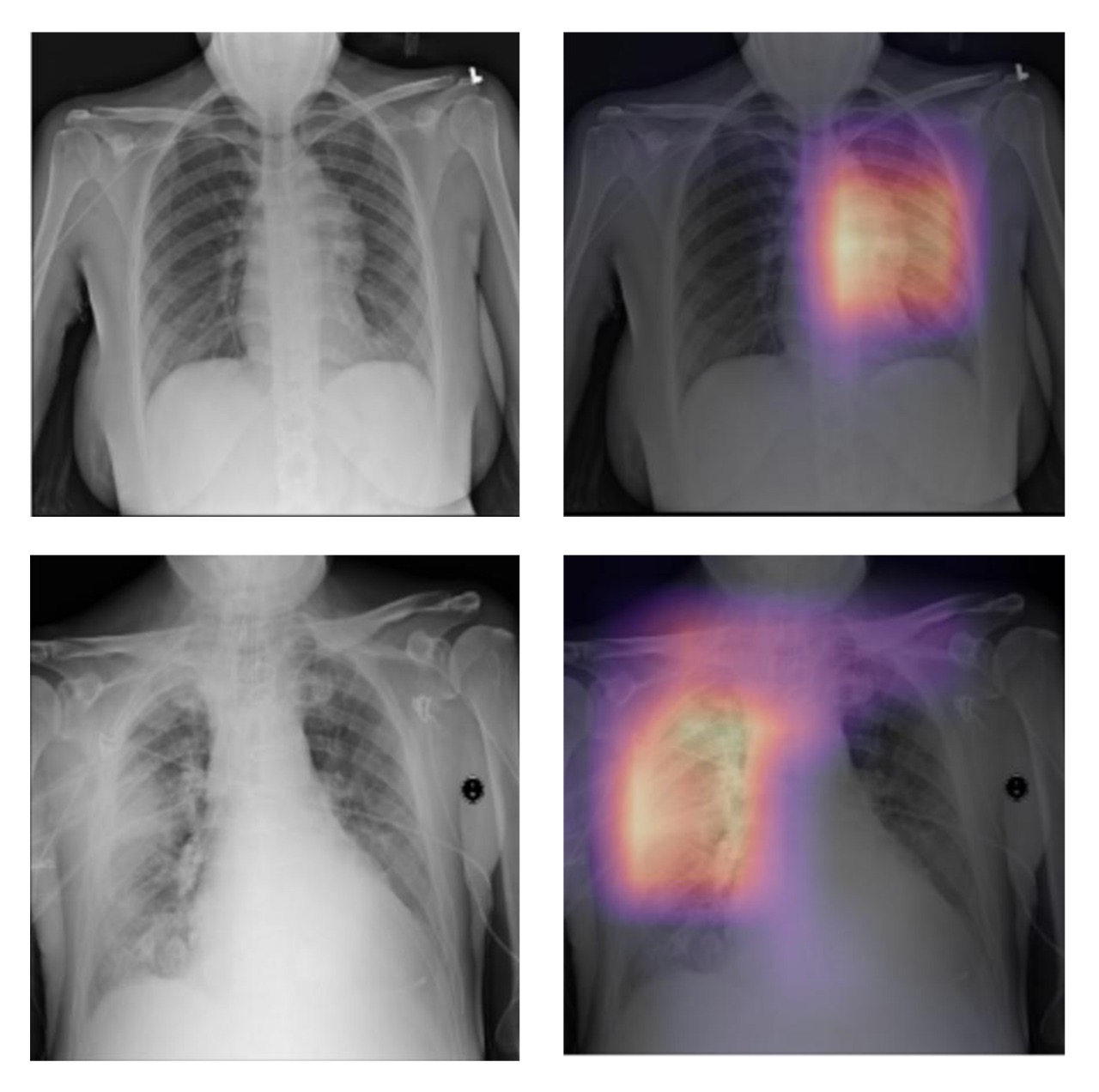}
\caption{A visualization of the attention  mechanism applied to Convolutional features overlaid on X-ray images }
\label{a4}
\end{figure}

\section{Related Work}
The outbreak of the COVID-19 [23] pandemic and the increasing count of the number of deaths have captured the attention of most researchers across the world.  Several works have been published which aim to either study this virus or in a way aim to curb the spread. Owing to the supremacy of computer vision and deep learning in the field of medical imaging, most of the researchers are using these tools as means to diagnose COVID-19. Chest X-ray (CXR) and Computed Tomography (CT) are the imaging techniques that play an important role in the detection of COVID-19 [24], [25].

As inferred from literature Convolutional Neural Network (CNN) remains the preferred choice of researchers for tackling COVID-19 from digitized images and several reviews have been carried out to highlight it’s recent contributions to COVID-19 detection[26]-[28]. For example in [29] a CNN based on Inception network was applied
to detect COVID-19 disease within computed tomography (CT). They achieved a total accuracy of 89.5\% with specificity of 0.88 and sensitivity of 0.87 on their internal validation and a total accuracy of 79.3\% with specificity of 0.83 and sensitivity of 0.67 on the external testing dataset. In [30]  a modified version of the ResNet-50 pre-trained network was used to classify CT images into three classes: healthy, COVID-19 and bacterial pneumonia. Their model results showed that the architecture could accurately identify the COVID-19 patients from others with an AUC of 0.99 and sensitivity of 0.93. Also their model could discriminate COVID-19 infected patients and bacteria pneumonia-infected patients with an AUC of 0.95, recall (sensitivity) of 0.96. In [31] a CNN architecture called COVID-Net based on transfer learning was applied to classify the Chest X- ray (CXR) images into four classes of normal, bacterial infection, non-COVID and COVID-19 viral infection. The architecture attained a best accuracy of 93.3\% on their test dataset. In [32] the authors proposed a deep learning model with 4 convolutional layers and 2 dense layers in addition to classical image augmentation and achieved 93.73\% testing accuracy. In [33] the authors presented a transfer learning method with a deep residual network for pediatric pneumonia diagnosis. The authors proposed a deep learning model with 49 convolutional layers and 2 dense layers and achieved 96.70\% testing accuracy. In [9] the authors proposed a modified CNN based on class decomposition, termed as Decompose Transfer Compose model to improve the performance of pre-trained models on the detection of COVID-19 cases from chest x-ray images. ImageNet pre-trained ResNet model was used for transfer-learning and they also used  Data Augmentation and histogram modification technique to enhance contrast of each image. Their proposed DeTraC-ResNet 18 model achieved an accuracy of 95.12\%. In [34] the authors proposed a  pneumonia chest x-ray detection based on generative adversarial networks (GAN) with a fine-tuned deep transfer learning for a limited dataset. The authors chose AlexNet, GoogLeNet, Squeeznet, and Resnet18 are selected as deep transfer learning models. The distinctive observation drawn for this paper was the use of GAN for generating similar examples of the dataset besides tackling the problem of overfitting. Their work used 10\% percent of the original dataset while generating the other 90\% using GAN. In [35] the authors presented a method to generate synthetic chest X-ray (CXR) images by developing an Auxiliary Classier Generative Adversarial Network (ACGAN) based model. Utilizing three publicly available datasets of IEEE Covid Chest X-ray dataset[10], COVID-19 Radiography Database [36] and COVID-19 Chest X-ray Dataset [37] the authors demonstrated that synthetic images produced by the ACGAN based model could improve the performance of CNN(VGG-16 in their case) for COVID-19 detection. The classification results showed an accuracy of 85\% with the CNN alone, and with the addition of the synthetically generated images via ACGAN the accuracy increased to 95\% .
Thus having understood the advantages that GAN offers on training models with relatively smaller datasets, in our research we implemented the DAGAN combined with the attention based Siamese Neural Networks for getting the optimum results out of a relatively smaller dataset used for training our model[10].

\section{Methods}
For our experiments the application was build using Android studio. MVVM (Model View View-Model) architecture has been used in the app, which helps in proper state management following the UI Material Design guidelines while building. For storage of the App data(like the local X-Rays samples) and Authentication, Firebase is used. The deep learning model was trained using publicly available datasets to test for robustness of the same. Some of the major issues with such datasets were lack of data, inherent noise features and class imbalance. In the proposed methodology, all three of these issues were tackled effectively. The smartphone application would pave way for improvement of the existing model and provide ease of access to state of the art disease diagnosis for common pulmonary diseases to everyone.A semi-live training scenario was build on the cloud which enables the model to imporve over time gradually without intervention.

\subsection{Application Usage Overview}
The android application acts as an accessible platform which assists doctors or patients in uploading the X-rays samples to be inferred by the deep learning model, and obtain corresponding diagnosis results, as seen in Figure 5. The application is as a cloud-user interface which enables wider accessibility and help the model imporve by the cloud build semi-live training scenario. The algorithm is deployed in a FAS ( Function as a Service ), which gets triggered when a user uploads a sample.
There are primarily 2 categories of users : doctors and patients. Users under the doctor category would be verified and could act as a potential source for labeled training data. Under the patient category the inference mechanism gets triggered which enables the backend model to provide the user with a diagnosis result for the uploaded sample.  
\begin{figure}[h!]
\includegraphics[width=12cm, height=10cm]{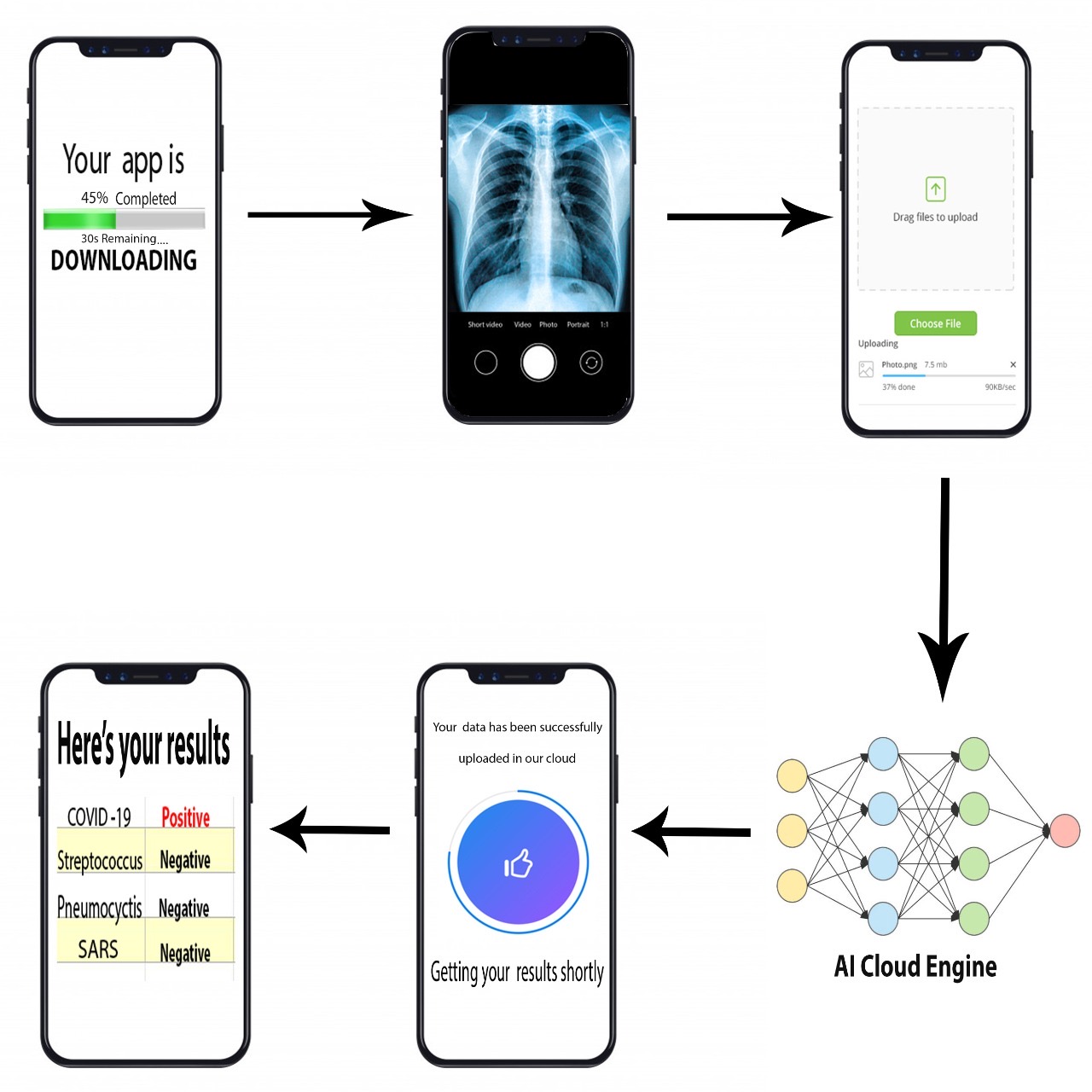}
\caption{Abstract workflow of our smartphone application utilizing an AI cloud platform }
\label{c1}
\end{figure}

\subsection{Backend Model}
The backend deep learning architecture mainly consists of two deep learning models- the DAGAN for robust and effective data augmentation, followed by the Convolutional Siamese Network with attention mechanism. The Siamese Network is proven to be data efficient through our experiments. Both of these networks are pretrained on publicly available datasets. To obtain the pretrained DAGAN model , suitably processed X-ray images were provided with corresponding class labels. For the pretrained Siamese Network,  visually variant augmented samples with in-class features preserved were generated using the DAGAN model. Then these generated samples were paired up for all possible combinations. Each of the pairs were assigned a binary label based on the classes on which the two images in a pair belonged to - 0 if both images are from the same class of pulmonary diseases and 1 otherwise. The resulting dataset was then used to train the Siamese Network. A set of well labelled and noise free images are selected to be the standard dataset for comparison. During inference procedure one of the twin among the Siamese Network generates a latent vector for the uploaded image by a forward pass. The second twin generates a latent vector for an image in the standard dataset. The obtained latent vectors are then compared using an energy function. The energy values of all classes in the standard dataset are obtained using a similar procedure, and the class with the lowest average value is selected. The class thus selected becomes the diagnosis for the particular uploaded image. The diagnosis made is conveyed back to the user through an online database.

\begin{figure}[h!]
\includegraphics[width=13cm, height=15cm]{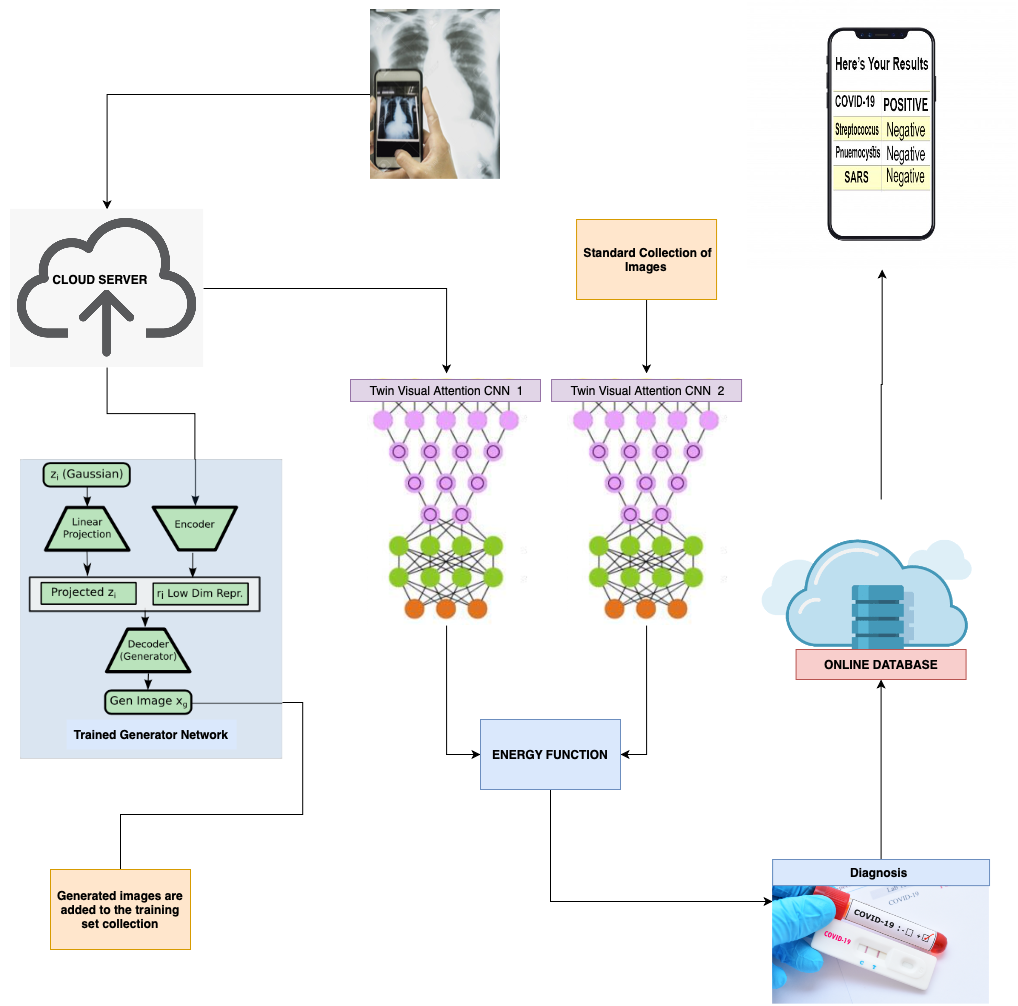}
\caption{General inference process for images}
\label{c2}
\end{figure}

\subsection{Semi-live training}
X-Ray images used by the backend model could show large variance due to a variety of reasons, which includes lighting condition while the picture is taken, the X-Ray machine specifications or camera quality of the user's smartphone etc. Since the challenges such as this due to data variation should be accounted in a real world scenario, a semi-live training scenario was introduced, which enables the model parameters of the pretrained model to further adapt to new or variant data. The scenario is triggered when sufficient amounts of data is obtained. 

\begin{figure}[h!]
\includegraphics[width=12cm, height=13cm]{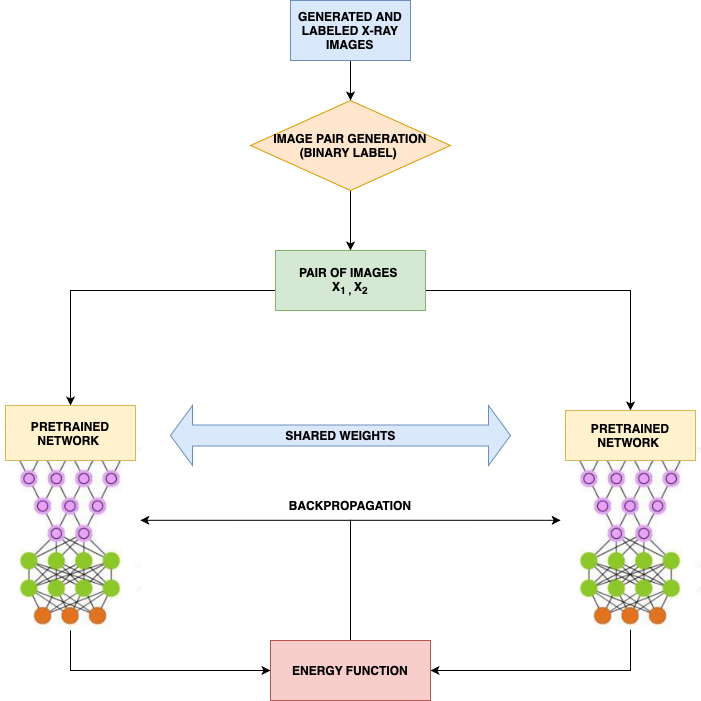}
\caption{Overview of the training process of the Siamese Network during semi-live training scenario}
\label{c3}
\end{figure}

\section{Experiments and Results}

Two different datasets are used to obtain a variety of comparison results for proper model evaluation. On the first and second datasets the tasks are formulated as binary and multiclass classification respectively. Dataset description and corresponding comparison results are given below. For the selection of the standard dataset, expert advice and third party help were utilized.
\subsection{Datasets}

\begin{figure}[h!]
\includegraphics[width=12cm, height=6cm]{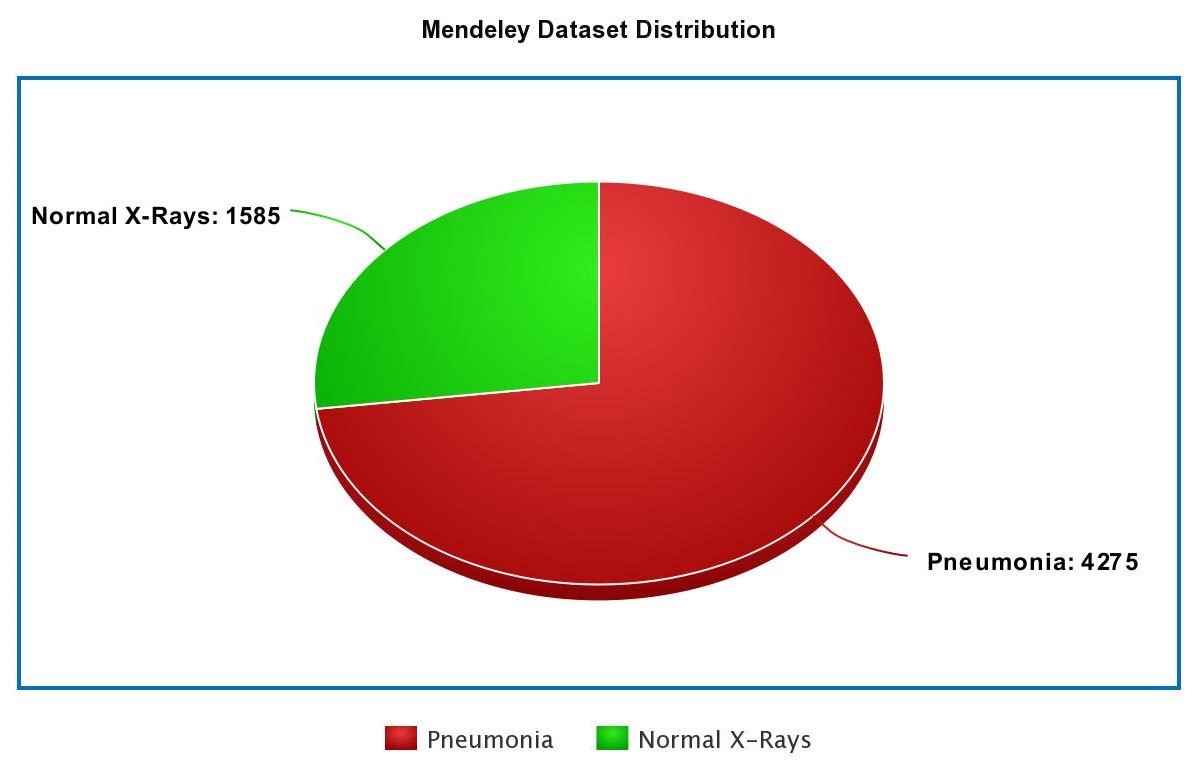}
\caption{Class distribution of dataset - 1 [10] (Graphical representation)}
\label{d1}
\end{figure}

\begin{figure}[h!]
\includegraphics[width=12cm, height=6cm]{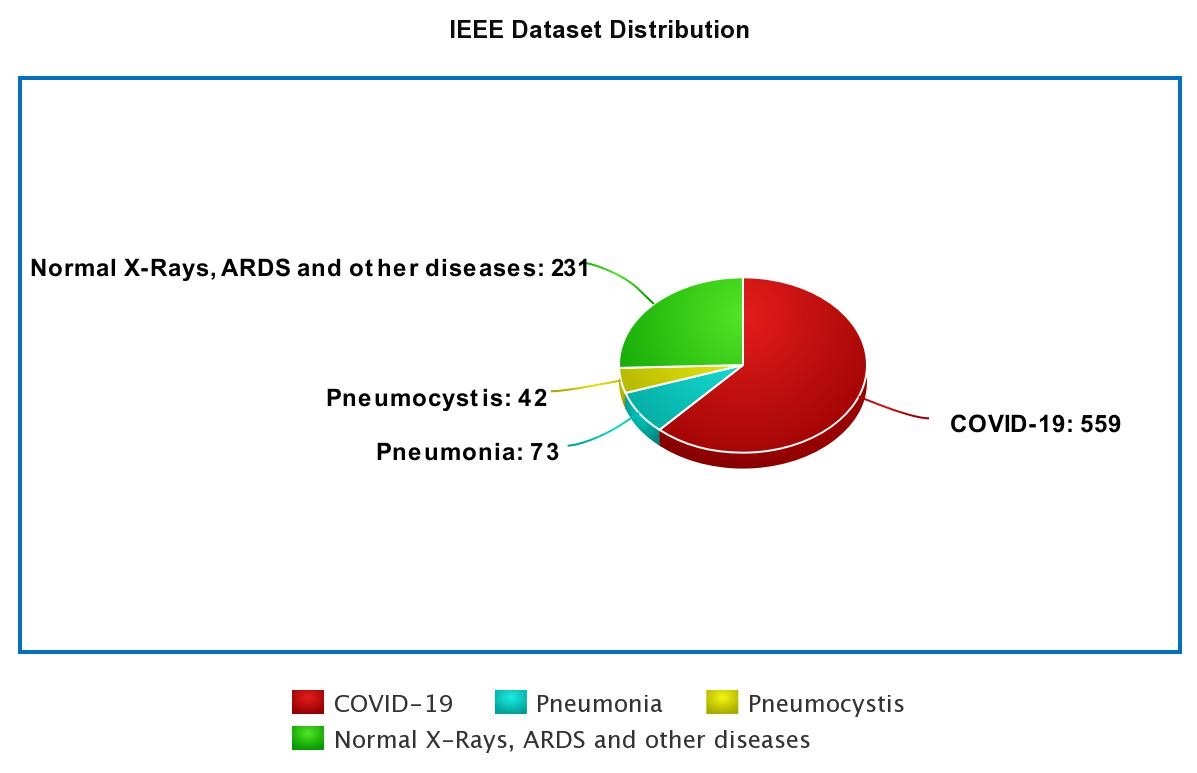}
\caption{Class distribution of dataset - 2 [38] (Graphical representation)}
\label{d2}
\end{figure}

Both the datasets used in this study are publicly available. Apart from the selection of the standard dataset, no specific dataset cleansing was done. Training process was done on the data including even those images with inherent noise features present, the same helps in confirming the robustness of the proposed model. Dataset-1 was published in 2018, the X-rays images obtained were part of clinical care conducted year to year from Guangzhou Medical Center from  5,863 different patients. Dataset-2 was published as an effort to give out relevant data for widespread studies that were conducted to tackle the COVID-19 pandemic situation. Test set size of datasets 1 and 2 were selected as 20\% of images from each class. The training set was further enlarged using the DAGAN model to ensure a generalized training for the proposed model.
\subsection{Comparative Study}

A good amount of images were selected as the testing set for the proposed model, so as to robustly test and evaluate the proposed method. As per the split of 20\%,  data from each class of the two datasets were randomly selected to be included in the test set. For dataset(1) used for the binary classification task the test set consisted of 1170 images out of the 5860 images, as for dataset(2) used for the multiclass classification task the test set consisted of 180 images out of the 905 images. No generation of images were done on the testing set, as it is considered important to conduct model evaluation on real world data samples.Since the testing set in both experiments are large, the confidence interval for the testing accuracy of the proposed model was calculated by assuming a Gaussian distribution for the dataset proportion.

\begin{table}[h]
\centering
\begin{tabular}{llll}
\hline
\textbf{Method} & \textbf{Year} & \textbf{Description}                                                                                                                      & \textbf{\begin{tabular}[c]{@{}l@{}}Testing \\ Accuracy\end{tabular}} \\ \hline
{[}39{]}         & 2018          & Convolutional Neural Network (CNN)                                                                                                        & 92.80\%                                                              \\ \hline
{[}40{]}         & 2019          & \begin{tabular}[c]{@{}l@{}}Deep learning model with 4 convolutional\\  layers and 2 dense layers + classical\\  Augmentation\end{tabular} & 93.73\%                                                              \\ \hline
{[}41{]}         & 2019          & \begin{tabular}[c]{@{}l@{}}Deep learning model with 7 convolutional\\  layers and 3 dense layers\end{tabular}                             & 95.30\%                                                              \\ \hline
{[}42{]}         & 2019          & \begin{tabular}[c]{@{}l@{}}Deep learning model with 49 convolutional\\  layers and 2 dense layers\end{tabular}                            & 96.70\%                                                              \\ \hline
{[}43{]}         & 2020          & \begin{tabular}[c]{@{}l@{}}Convolutional Neural Network (CNN) \\ + Random forest\end{tabular}                                             & 97.00\%                                                              \\ \hline
{[}34{]}         & 2020          & GAN + Resnet18                                                                                                                            & 99.00\%                                                              \\ \hline
\begin{tabular}[c]{@{}l@{}}Proposed \\ Method\end{tabular} & 2020          & DAGAN + Attention Siamese Net                                                                                                             & 99.30 +/- 0.63\%                                                              \\ \hline
\end{tabular}
\caption{Comparison of testing accuracy of proposed model with related works conducted on dataset-1}
\label{tab:my-table}
\end{table}

\begin{table}[h]
\centering
\begin{tabular}{llll}
\hline
\textbf{Method}                                            & \textbf{Year} & \textbf{Description}                                                                                         & \textbf{\begin{tabular}[c]{@{}l@{}}Testing \\ Accuracy\end{tabular}} \\ \hline
{[}44{]}                                                    & 2020          & Using pre-trained model of CheXNet                                                                           & 90.50\%                                                              \\ \hline
{[}45{]}                                                    & 2020          & \begin{tabular}[c]{@{}l@{}}Extracts the features from chest x-ray \\ images using FrMEMs moment\end{tabular} & 96.09\%                                                              \\ \hline
{[}46{]}                                                    & 2020          & \begin{tabular}[c]{@{}l@{}}Two-level Hierarchical Deep Neural \\ Network and transfer learning\end{tabular}  & 97.80\%                                                              \\ \hline
\begin{tabular}[c]{@{}l@{}}Proposed \\ Method\end{tabular} & 2020          & DAGAN + Attention Siamese Net                                                                                & 98.40 +/- 2.18\%                                                              \\ \hline
\end{tabular}
\caption{Comparison of testing accuracy of proposed model with related works conducted on dataset-2}
\label{tab:my-table 1}
\end{table}

The tables illustrate the robustness of the proposed model as well as point out how effective the model is under the scenario, where there is a lack of availability of training data.
\subsection{Final Model Analysis}

For our purposes, in order obtain a robust and deployable model, we combine both datasets and train a multiclass classification model which is robustly evaluated for performance. The testing set is selected to be 20\% of images from each class, at random. The model thus obtained, achieved a testing accuracy of 97.8\%. The validation set is selected to be 20\% of images in each class, from the training set.%

\begin{figure}[h!]
\includegraphics[width=12cm, height=6cm]{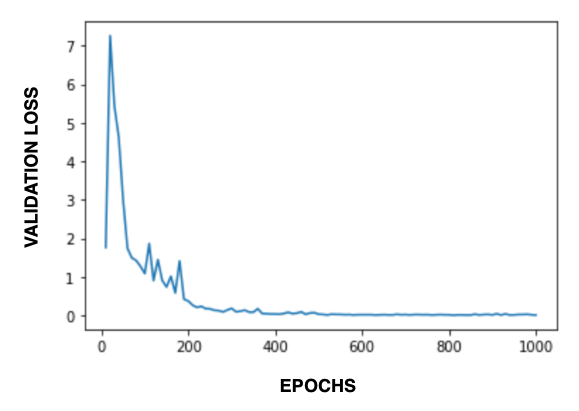}
\caption{The validation loss vs epoch curve}
\label{co1}
\end{figure}

\begin{figure}[h!]
\includegraphics[width=12cm, height=7cm]{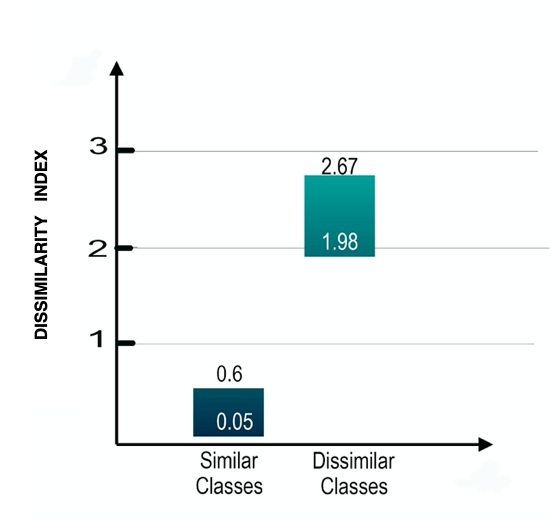}
\caption{Range of variation of dissimilarity index for like and unlike classes}
\label{co2}
\end{figure}

Detailed analysis was carried to determine the class separation boundary of the latent space, Figure 11 illustrates the large class separation found. 

The illustration (Figures [12-14]) shows how effective is the latent space representation so formed by training the model, in representing the lower dimensional projection of images.
\begin{figure}[h!]
\includegraphics[width=12cm, height=7cm]{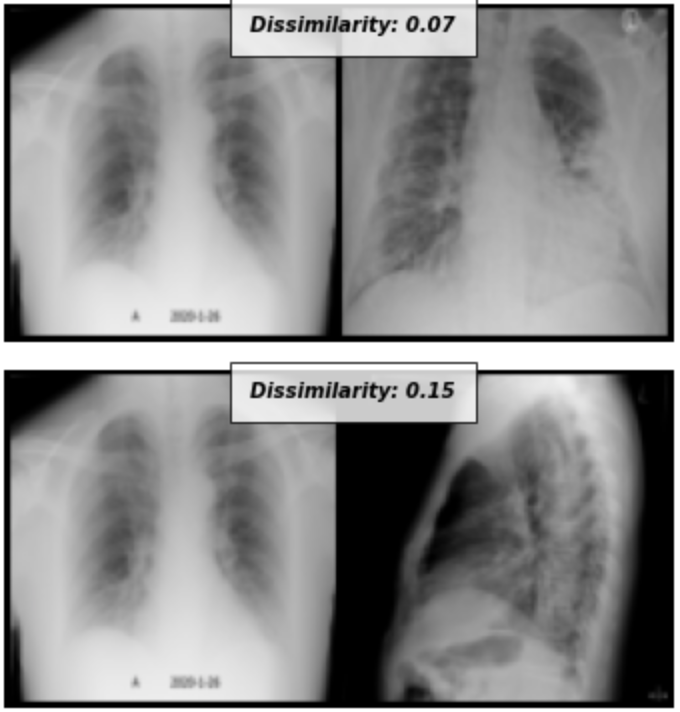}
\caption{Comparing a chest X-Ray image of a COVID-19 positive patient with test set images in the COVID-19 class }
\label{co3}
\end{figure}

\newpage

\begin{figure}[h!]
\includegraphics[width=12cm, height=7cm]{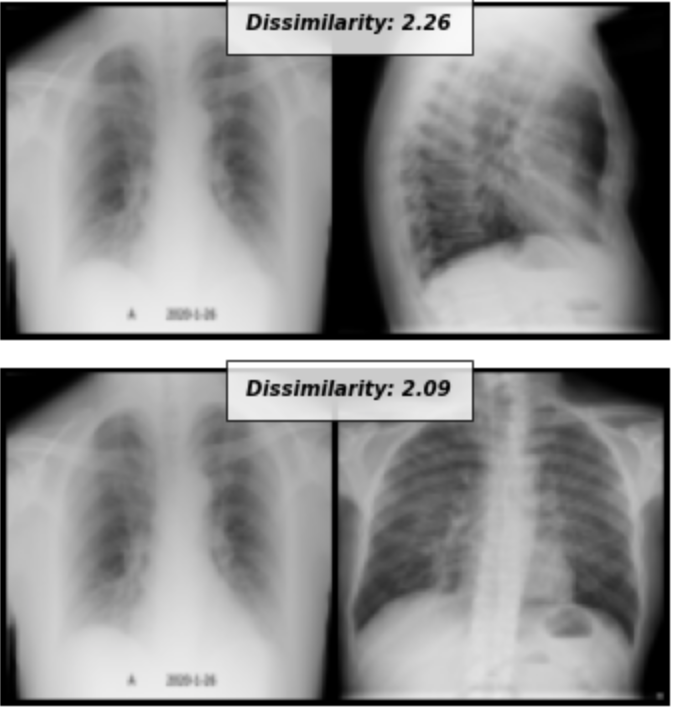}
\caption{Comparing a chest X-Ray image of a COVID-19 positive patient with test set images in the Pneumocystis class}
\label{co4}
\end{figure}

\begin{figure}[h!]
\includegraphics[width=12cm, height=7cm]{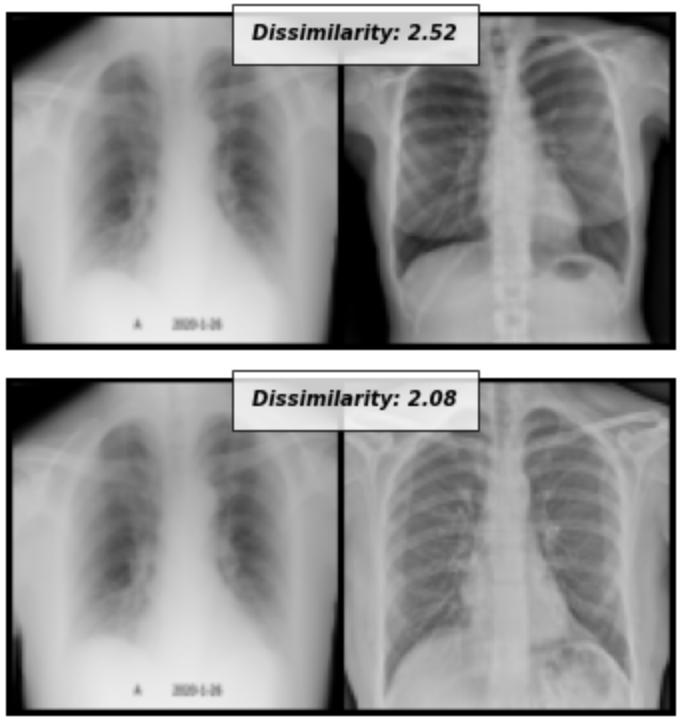}
\caption{Comparing a chest X-Ray image of a COVID-19 positive patient with test set images in the Normal class}
\label{co5}
\end{figure}

The dissimilarity index values are high when unlike classes are compared, at the same time a low dissimilarity is obtained when like classes are compared. 

\section*{Conclusion}
In the wake of the global pandemic preventive and therapeutic solutions are in the limelight as doctors and healthcare professionals work to tackle the threat, with diagnostic methods having extensive capabilities being the need of the hour. The COVID-19 outbreak has caused an adverse effect in all walks of day to day life worldwide. Fact remains that the spread of such a disease could have been prevented during the early stages, with the help of  accurate methods of diagnosis. Medical images such as X-rays and CT scans are of great use when it comes to disease diagnosis, particularly chest X-rays being pivotal in diagnosis of many common and dangerous respiratory diseases. Radiologists can infer many crucial facts from a chest X-ray which can be put to use in diagnosing several diseases. Today’s AI methods that mimic disease diagnosis as done by radiologists could outperform any human radiologist, owing to the higher pattern recognition capabilities and the lack of the human element of error or inefficiency in turn paving way for extensive research in this area. A common and efficient method to employ would be to use a Convolutional Neural Network (CNN) based classifier , which could accurately recognise patterns from images to make necessary predictions. The limitation of the same being the requirement of huge amounts of data to obtain a classifier model with enough generalizability and accuracy. Metrics improvement of an existing model is a hard task since retraining process for a large deep learning model would be expensive in terms of time and computation, vulnerable to scalability issues in retraining. Hence we adopted feature comparison based methods which are superior to feature recognising methods in these respects, exploiting a deep Generative Network for data augmentation. The model exhibited profound comparison metrics having very distinguishable dissimilarity indices. Similar classes showed remarkably low indices ranging from 0.05 to 0.6 , while different classes had higher values lying between 1.98 and 2.67. These performance indices of dissimilarity and the the large gap between these classes consolidates the fact that our model is able to clearly demarcate and classify diseases with state of the art efficiency.

The limitations of this study include the inevitable noise factor on the dataset used, to tackle the same a cloud based live training method has been employed which uses properly annotated and identified data from medical practitioners worldwide. The underlying method could be employed to detect several other diseases if necessary, modification required for the current model minimal as compared to any deep learning based backend systems. Doctors and Radiologists can leverage the ability of our application to make a reliable remote diagnosis, thereby saving considerable time which can be devoted to medication or prescriptive measures. Due to the high generalisability and data efficiency of the method , the application could prove itself to be a great tool in not only in accurately diagnosing diseases of interest, but to also conduct crucial studies on emerging or rare respiratory conditions.

\section*{Code Availability}
The custom Python code and android app used in this study are available from the corresponding author upon reasonable request and is to be used only for
educational and research purposes.

\end{document}